\documentclass[aps,prl,reprint,showpacs,superscriptaddress,twocolumn,10pt]{revtex4-1}%
\usepackage{amsfonts}
\usepackage{mathrsfs}
\usepackage{amsmath}
\usepackage{amssymb}
\usepackage{graphicx}
\usepackage{color}%
\usepackage{mathtools}
\usepackage{booktabs}
\usepackage{mleftright}
\usepackage[colorlinks,linkcolor=blue,citecolor=blue,urlcolor=blue,hyperindex,bookmarks=false,pdfstartview=FitH]{hyperref}

\providecommand{\U}[1]{\protect\rule{.1in}{.1in}}
\newcommand{\blue}[1]{\textcolor[rgb]{0.00,0.00,0.80}{#1}}

\newcommand{\figpanel}[2]{\hyperref[#1]{\ref*{#1}(#2)}}

\begin{document}
\title{Giant-Atom Effects on Population and Entanglement Dynamics of Rydberg Atoms}
\author{Yao-Tong Chen}
\affiliation{School of Physics and Center for Quantum Sciences,
Northeast Normal University, Changchun 130024, China}
\author{Lei Du}
\email{dul256@nenu.edu.cn} \affiliation{School of Physics and Center
for Quantum Sciences, Northeast Normal University, Changchun 130024,
China}
\author{Yan Zhang}
\affiliation{School of Physics and Center for Quantum Sciences,
Northeast Normal University, Changchun 130024, China}

\author{Lingzhen Guo}
\affiliation{Center for Joint Quantum Studies and Department of
Physics, School of Science, Tianjin University, Tianjin 300072,
China}

\author{Jin-Hui Wu}
\email{jhwu@nenu.edu.cn} \affiliation{School of Physics and Center
for Quantum Sciences, Northeast Normal University, Changchun 130024,
China}
\author{M. Artoni}
\affiliation{Department of Chemistry and Physics of Materials,
University of Brescia, Italy} \affiliation{European Laboratory for
Non-Linear Spectroscopy, Sesto Fiorentino, Italy}
\author{G. C. La Rocca}
\affiliation{NEST, Scuola Normale Superiore, Piazza dei Cavalieri 7,
I-56126 Pisa, Italy}

\date{\today }

\begin{abstract}
Giant atoms are attracting interest as an emerging paradigm in the
quantum optics of engineered waveguides. Here we propose to realize
a synthetic giant atom working in the optical regime starting from a
pair of interacting Rydberg atoms driven by a coherent field and
coupled to a photonic crystal waveguide. Giant-atom effects can be
observed as a phase-dependent decay of the double Rydberg excitation
during the initial evolution of this atomic pair while (internal)
atomic entanglement is exhibited at later times. Such an intriguing
entanglement onset occurs in the presence of intrinsic atomic decay
toward non-guided vacuum modes and is accompanied by an
anti-bunching correlation of the emitted photons. Our findings may
be relevant to quantum information processing, besides broadening
the giant-atom waveguide physics with optically driven natural
atoms.

\end{abstract}

\maketitle

\textit{Introduction}.---A fascinating paradigm of quantum optics
dubbed ``giant atom'' has been developed recently to describe
situations where artificial atoms interact with microwave or
acoustic fields beyond the standard dipole
approximation~\cite{review}. With state-of-the-art technologies, one
can implement giant atoms that are coupled to a waveguide at
multiple points, with coupling separations comparable to field
wavelengths. These nonlocal interactions can result in peculiar
self-interference effects, which account for a number of phenomena
not happening in conventional natural atoms, such as
frequency-dependent atomic relaxation rates and Lamb
shifts~\cite{2014,2020nature,continuum1,cp2022,liao}, non-exponential atomic
decay~\cite{decay1,decay2,decay3}, in-band decoherence-free
interactions~\cite{2020nature,free,fra,AFKchiral}, and exotic
atom-photon bound states~\cite{bound1,bound2,bound3,bound4,bound5,bound6}.
To date, platforms capable of implementing giant atoms mainly
include superconducting quantum
circuits~\cite{saw14,2020nature,decay2,Wilson2021}, coupled
waveguide arrays~\cite{longhiretard}, and cold atoms in
state-dependent lattices~\cite{2019}, but have also been extended to
synthetic frequency dimensions~\cite{DLprl} and ferromagnetic spin
systems~\cite{fer}.

Nevertheless, it is important to explore new physics of giant atoms
within different atomic architectures, especially those working
beyond the microwave or acoustic regime and with high-lying Rydberg
atoms. Owing to unique properties unlike atoms in the ground or
low-lying excited states, including strong dipole-dipole
interactions and long radiative lifetimes~\cite{ryd2010}, Rydberg
atoms turn out to be an excellent building block in quantum
information processing, with potential applications for realizing
quantum logic gates~\cite{gate1,gate2,gate3,gate4}, single-photon
sources~\cite{source1,source2,source3,source4}, and various
entangled states~\cite{en1,en2,en3,en4}. Rydberg atoms have also
been successfully employed in waveguide quantum electrodynamics by
coupled to engineered structures like optical
nanofibers~\cite{nano}, photonic crystal waveguides~\cite{pcw}, and
coplanar microwave waveguides~\cite{cop}.

In this Letter, we consider a pair of two-level Rydberg atoms
coupled to a photonic crystal waveguide (PCW) through two lower
transitions and driven by a coherent field through two upper
transitions in the two-atom basis to display giant-atom effects in
the optical regime. The self-interference effect appears
specifically at \textit{shorter times} of the pair's dynamic
evolution dominated by two competing two-photon resonant
transitions, thereby realizing a \textit{synthetic} giant atom with
two coupling points at a variable distance $d$ about the atomic
separation $R$. For \textit{longer times}, on the other hand, we
observe the onset of atomic entanglement through mutual Rydberg
excitations of the two atoms as the phase $\phi$ accumulated from a
coupling point to the other takes specific values, which depends on
$d$ and is tunable. This effect is characterized by a detailed
examination on \textit{quantum correlations} of the emitted photons
and further understood in terms of \textit{dark states} decoupled
from both coherent field and waveguide modes. Our findings remain
valid even for continuous atom-waveguide couplings, which are more
appropriate for Rydberg atoms with large spatial extents.

\vspace{1mm}

\begin{figure}[ptb]
\centering
\includegraphics[width=8.5 cm]{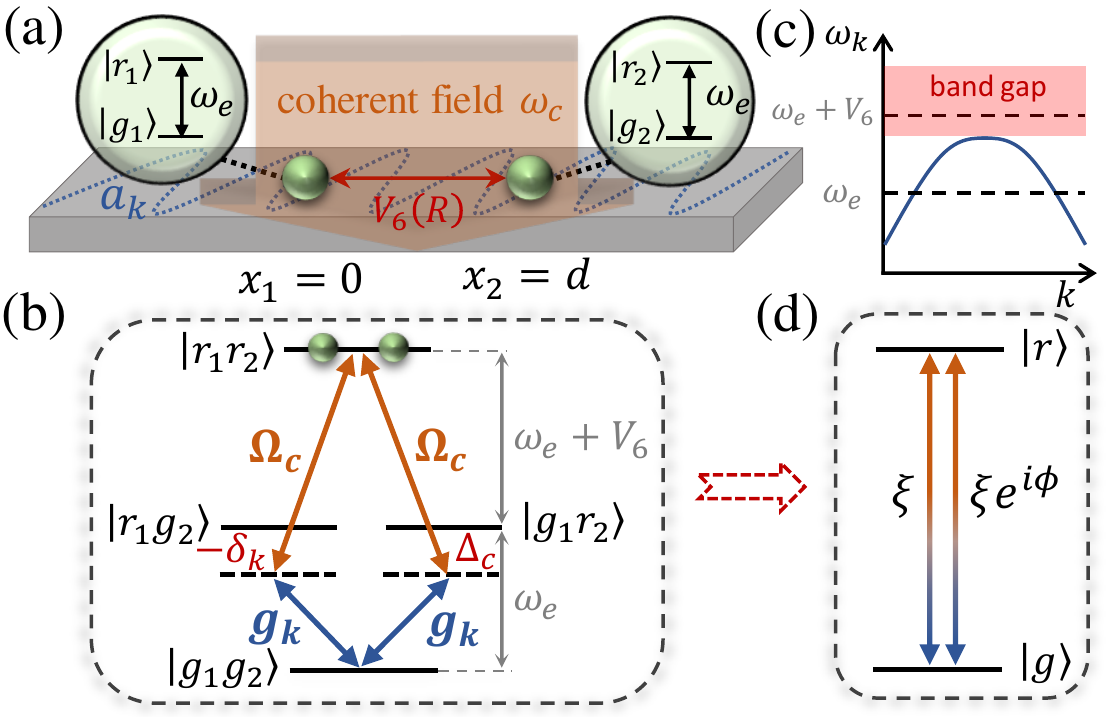}
\caption{(a) The two-level configuration (\textit{single-atom
basis}). Two Rydberg atoms placed at a distance $R$ interact via a
potential $V_{6}(R)$ and couple to a waveguide mode $a_{k}$ at
$x_{1}=0$ or $x_{2}=d$ while driven by a coherent field
$\omega_{c}$. (b) The four-level configuration (\textit{two-atom
basis}). Blue (red) lines represent single (double) Rydberg
excitations, whereby two lower (upper) transitions are coupled by
mode $a_{k}$ of strength $g_{k}$ (field $\omega_{c}$ of strength
$\Omega_{c}$). (c) The dispersion relation of a PCW: frequency
$\omega_e$ of the lower transitions falls within the propagating
band while frequency $\omega_e+V_6$ of the upper transitions falls
within the band gap (red shaded). (d) An equivalent two-level giant
atom with $|g\rangle\equiv|g_{1}g_{2}\rangle$ and
$|r\rangle\equiv|r_{1}r_{2}\rangle$, upon adiabatic elimination of
the single-excitation states, bears separate couplings of strengths
$\xi$ at $x_{1}$ and $\xi e^{i\phi}$ at $x_{2}$, with $\phi$ being
the accumulated phase between $x_1$ and $x_2$. } \label{fig1}
\end{figure}

\textit{Model}.--We start by illustrating in Fig.~\figpanel{fig1}{a}
that a pair of Rydberg atoms with ground $|g_{1,2}\rangle$ and
Rydberg $|r_{1,2}\rangle$ states separated by frequency $\omega_{e}$
are trapped in the vicinity of a PCW~\cite{pcw1,pcw2} at $x_{1}=0$ and
$x_{2}=d$, respectively. The two atoms are illuminated by a coherent
field of frequency $\omega_{c}$ and interact through a (repulsive)
van der Waals (vdW) potential $V_{6}=C_{6}/R^{6}$, when
\textit{both} excited to the Rydberg states, with $C_{6}$ being the
vdW coefficient and $R$ the interatomic distance, which could be
different from $d$~\cite{supp}. The vdW interaction will result in a
largely shifted energy $2 \hbar \omega_{e}+\hbar V_{6}$ for the
\textit{double-excitation} state $|r_{1}r_{2}\rangle$ while the
\textit{single-excitation} states $|r_{1}g_{2}\rangle$ and
$|g_{1}r_{2}\rangle$ remain to exhibit energy $\hbar\omega_{e}$. In
the two-atom basis, for a large enough $V_{6}$, we have the
four-level configuration shown in Fig.~\figpanel{fig1}{b} whereby
the coherent field $\omega_c$ drives only transitions
$|g_{1}r_{2}\rangle\leftrightarrow|r_{1}r_{2} \rangle$ and
$|r_{1}g_{2}\rangle\leftrightarrow|r_{1}r_{2} \rangle$ with detuning
$\Delta_{c}=\omega_{c}-(\omega_{e}+V_{6})$ and strength
$\Omega_{c}$, while a waveguide mode of frequency $\omega_{k}$
(wavevector $k$) drives only transitions
$|g_{1}g_{2}\rangle\leftrightarrow|g_{1}r_{2}\rangle$ and
$|g_{1}g_{2}\rangle\leftrightarrow|r_{1}g_{2}\rangle$ with detuning
$\delta_{k}=\omega_{k}-\omega_{e}$ and strength $g_{k}$. This scheme
is supported by the following two considerations. \textit{First},
$\omega_{c}$ is far away from $\omega_{e}$ but close to
$\omega_{e}+V_{6}$ with $\Delta_{c}\ll V_{6}$, so that the coherent
field can only drive two upper transitions. \textit{Second},
$\omega_{e}$ and $\omega_{e}+V_{6}$ fall, respectively, within the
lower propagating band and the band gap of a PCW as sketched in
Fig.~\figpanel{fig1}{c}, so that the waveguide mode can only drive
two lower transitions.

\vspace{1mm}

Then, for a continuum of waveguide modes interacting only with two
lower transitions, the Hamiltonian in the rotating-wave
approximation is ($\hbar=1$)
\begin{equation}
\begin{split}
H&=\omega_{e}(\sigma_{+}^{1}\sigma_{-}^{1}+\sigma_{+}^{2}\sigma_{-}^{2})+(\omega_{e}+V_{6}/2)(\sigma_{+}^{3}\sigma_{-}^{3}+\sigma_{+}^{4}\sigma_{-}^{4})\\
&\quad\,+\int dk\omega_{k}a_{k}^{\dagger}a_{k}+\Big[\int dkga_{k}\big(\sigma_{+}^{1}+e^{ikd}\sigma_{+}^{2}\big)\\
&\quad\,+\Omega_c
e^{-i\omega_{c}t}(\sigma_{+}^{3}+\sigma_{+}^{4})+\text{H.c.}\Big].
\label{H4}
\end{split}
\end{equation}
Here, we have introduced the atomic raising operators
$\sigma_{+}^{1}=|r_{1}g_{2} \rangle\langle g_{1}g_{2}|$,
$\sigma_{+}^{2}=|g_{1}r_{2} \rangle\langle g_{1}g_{2}|$,
$\sigma_{+}^{3}=|r_{1}r_{2} \rangle\langle r_{1}g_{2}|$, and
$\sigma_{+}^{4}=|r_{1}r_{2} \rangle\langle g_{1}r_{2}|$  (two-atom
basis) while the corresponding lowering operators are
$\sigma_{-}^{j}=(\sigma_{+}^{j})^{\dagger}$. Moreover,
$a_{k}^{\dagger}$ and $a_{k}$ refer, respectively, to the creation
and annihilation operators of a waveguide mode $\omega_{k}$. We have
also assumeed constant coupling strengths, i.e. $g_{k}\simeq g$, in
the Weisskopf-Wigner approximation.

\vspace{1mm}

As we address only the two-atom dynamics, the waveguide modes of
density $D(k)$ at frequency $\omega_{k}$ can be traced out
(Born-Markov approximation), yielding the master equation for
density operator $\rho$~\cite{book1,20161,supp}
\begin{equation}
\begin{split}
\partial_{t}\rho&=-i[H_{at},\rho]+\sum_{i=1}^{4}\gamma\mathcal{L}[\sigma_{-}^{i}]\rho+\sum_{j=1}^{2}\Gamma\mathcal{L}[\sigma_{-}^{j}]\rho\\
&\quad\,+\Gamma_{ex}\left[(\sigma_{-}^{1}\rho\sigma_{+}^{2}-\frac{1}{2}\{\sigma_{+}^{1}\sigma_{-}^{2},\rho\})+\text{H.c.}\right],\\
\label{master4}
\end{split}
\end{equation}
where $\mathcal{L}[O]\rho=O\rho
O^{\dagger}-\frac{1}{2}\{O^{\dagger}O,\rho\}$ is the Lindblad
superoperator describing two-atom decay processes, with
$\Gamma(k)=4\pi g^{2}D(k)$ denoting the decay rate into relevant
modes of the waveguide (guided modes)~\cite{2014,2020nature}
while $\gamma$ being the mean decay rate into other electromagnetic
modes (non-guided modes). Under the two-photon resonance condition
$\Delta_{c}+\delta_{k}\simeq0$, assumed to hold for all guided
modes, the atomic Hamiltonian is
\begin{equation}
\begin{split}
H_{at}&=\Delta_{c}(\sigma_{+}^{1}\sigma_{-}^{1}+\sigma_{+}^{2}\sigma_{-}^{2})+J_{ex}(\sigma_{+}^{1}\sigma_{-}^{2}+\sigma_{+}^{2}\sigma_{-}^{1})/2\\
&\quad\,+[\Omega_{c}(\sigma_{+}^{3}+\sigma_{+}^{4})+\text{H.c.}],
\label{Hat}
\end{split}
\end{equation}
with $J_{ex}=\Gamma\text{sin}\phi$ and
$\Gamma_{ex}=\Gamma\text{cos}\phi$ denoting, respectively,
\emph{coherent} and \emph{dissipative} parts of the exchange
interaction mediated by the waveguide. Here we have defined
$\phi(k)=|k|d=\omega_{k}d/|v_{g}(k)|\simeq\omega_{e}d/|v_{g}(k)|$ with
$v_{g}(k)$ being the group velocity of a guided mode bearing the
``linearized" dispersion $\omega_{k}\simeq k v_{g}(k)$. It is worth
stressing that $\Gamma_{ex}$ serves as a reservoir to
\textit{engineer} the atomic decay through selected guided modes
$\omega_k$ (or bandwidth of modes) and their density distribution
$D(k)$. In turn, such an engineered reservoir together with the
atomic separation $d$ represents a set of knobs to control the phase
$\phi(k)$ acquired by an emitted photon between the \textit{contact
points} $x_1$ and $x_2$ of a giant-atom (see below).

\begin{figure}[ptb]
\centering
\includegraphics[width=8.5 cm]{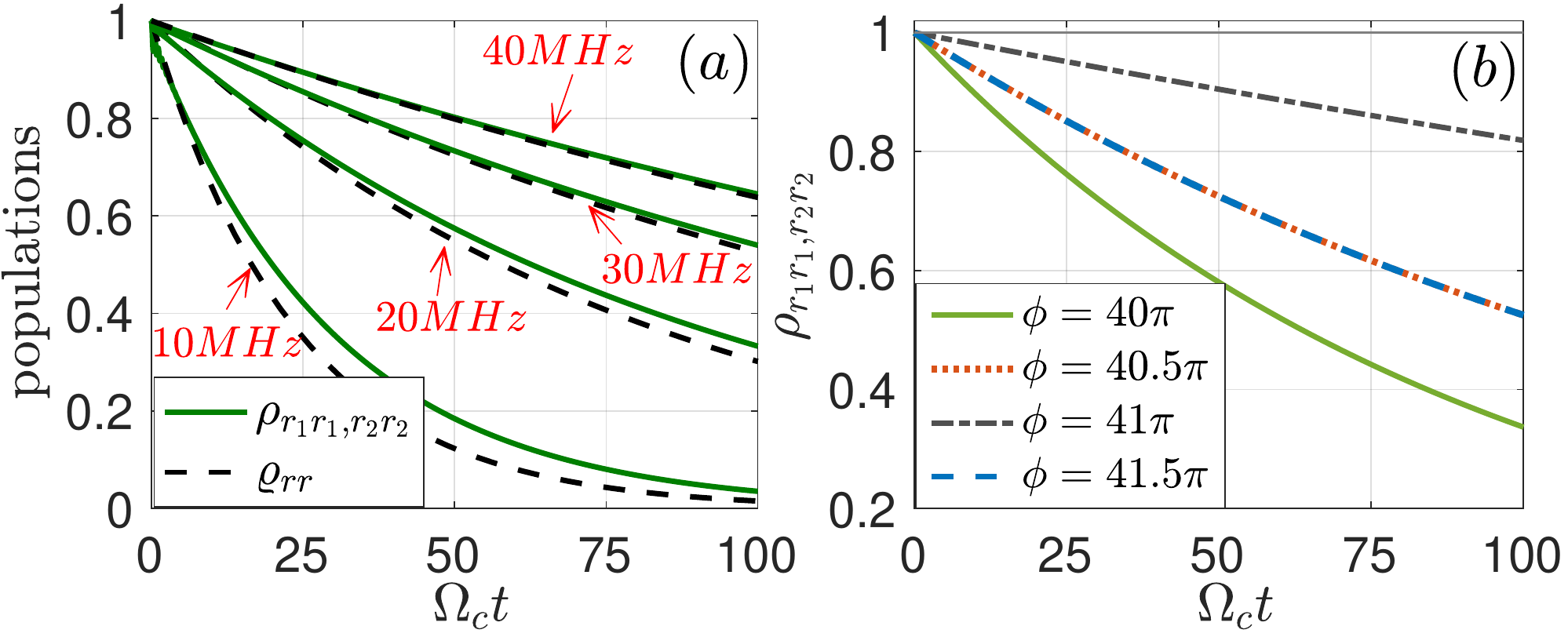}
\caption{(a) Time evolutions of double Rydberg population
$\rho_{r_{1}r_{1},r_{2}r_{2}}$ and giant-atom population
$\varrho_{rr}$ with $\phi=40.5\pi$ and different values of
$\Delta_{c}$. (b) Time evolutions of $\rho_{r_{1}r_{1},r_{2}r_{2}}$
with $\Delta_{c}=30$ MHz and different values of $\phi$. Other
parameters are $V_{6}=20$ GHz, $\Omega_{c}=1.0$ MHz, $\Gamma=1.0$
MHz, and $\gamma=1.0$ kHz. The thin gray line in (b) is shown as a
reference for $\gamma=0$ (no intrinsic decay) with $\Delta_{c}=30$
MHz and $\phi=41\pi$.} \label{fig2}
\end{figure}

\vspace{1mm}

\textit{Synthetic two-level giant atom: the short-time regime}.-
Maintaining $\Delta_{c}+\delta_{k}\simeq0$ and further requiring
$|\Delta_{c}|\gg\Omega_{c},g$, the two atoms initially in the double
Rydberg state $|r_{1}r_{2}\rangle$ would behave like a two-level
giant atom decaying directly to the ground state
$|g_{1}g_{2}\rangle$ at two coupling points $x_1$ and $x_2$. This
expectation will be verified by numerically comparing the dynamics
of the four-level atomic pair to that of the synthetic two-level
giant atom. In the latter picture, the two atoms interact with the
waveguide modes only through the two-photon resonant transition
$|r_{1}r_{2}\rangle\leftrightarrow|g_{1}g_{2}\rangle$ whereby an
external photon of frequency $\omega_c$ and a waveguide photon of
frequency $\omega_k$ are emitted (or absorbed) at the same time.
Upon the adiabatic elimination of states $|r_{1}g_{2}\rangle$ and
$|g_{1}r_{2}\rangle$~\cite{elimination1,elimination2}, the effective
$|r_{1}r_{2}\rangle\leftrightarrow|g_{1}g_{2}\rangle$ transition
amplitude consists of two contributions:
$\xi_1=-g\Omega_{c}/\Delta_{c}\equiv\xi$ (interaction at $x_{1}=0$)
and $\xi_2=\xi e^{i\phi}$ (interaction at $x_{2}=d$), as sketched in
Fig.~\figpanel{fig1}{d}, which interfere with each other.

\vspace{1mm}

With the above assumptions, we write down the synthetic two-level
giant-atom Hamiltonian as
\begin{equation}
\begin{split}
\mathcal{H}&=(2\omega_{e}+V_{6})\sigma_{+}\sigma_{-}+\int dk(\omega_{k}+\omega_{c})a_{k}^{\dagger}a_{k}\\
&\quad\,+\int dk\big[\xi
a_{k}\big(1+e^{ikd}\big)\sigma_{+}+\text{H.c.}\big] \; ,
\label{Heff}
\end{split}
\end{equation}
in terms of the transition operator
$\sigma_{+}=(\sigma_{-})^{\dagger}=|r\rangle\langle g|$ with
$|r\rangle\equiv|r_{1}r_{2}\rangle$ and
$|g\rangle\equiv|g_{1}g_{2}\rangle$. Again, by tracing out the
waveguide modes, we arrive at the master equation for giant-atom
density operator $\varrho$~\cite{supp}
\begin{equation}
\begin{split}
\partial_{t}\varrho&=2\gamma\mathcal{L}[\sigma_{-}]\varrho+\big(\Upsilon+\Upsilon^{\ast}\big)\sigma_{-}\varrho\sigma_{+}\\
&\quad\,\,-\Upsilon\sigma_{+}\sigma_{-}\varrho-\Upsilon^{\ast}\varrho\sigma_{+}\sigma_{-},
\label{master2}
\end{split}
\end{equation}
where
$\Upsilon=4\pi\xi^{2}D(1+e^{i\phi})=(\Gamma+\Gamma_{ex}+iJ_{ex})\Omega_{c}^{2}/\Delta_{c}^{2}$
with its real and imaginary parts being, respectively, the
phase-dependent decay rate and Lamb shift.

\vspace{1mm}

In the remaining part, we perform numerical analysis in support of
the predictions anticipated above. We first consider that for large
enough (driving) detunings $|\Delta_{c}|$, the adiabatic elimination
of two single-excitation states can be actually made, providing in
turn an adequate evidence of the equivalence between the
(four-level) atomic pair and the (two-level) giant atom in the
short-time regime. This has been examined in Fig.~\figpanel{fig2}{a}
by comparing time evolutions of atomic-pair population
$\rho_{r_{1}r_{1},r_{2}r_{2}}$ based on Eq.~(\ref{master4}) and
giant-atom population $\varrho_{rr}$ based on Eq.~(\ref{master2})
with matched parameters. Taking $\phi=40.5\pi$ as an example and
starting from $\rho_{r_{1}r_{1},r_{2}r_{2}}(0)=\varrho_{rr}(0)=1$,
we find that $\rho_{r_{1}r_{1},r_{2}r_{2}}(t)$ and $\varrho_{rr}(t)$
exhibit a better agreement for a larger $|\Delta_{c}|$ so that the
adiabatic elimination leading to a giant atom becomes reliable for
$|\Delta_{c}|/\Omega_{c}\gtrsim 30$. It is also worth noting that
$\rho_{r_{1}r_{1},r_{2}r_{2}}(t)$ and $\varrho_{rr}(t)$ decay faster
as $|\Delta_{c}|$ decreases because a smaller $|\Delta_{c}|$ results
in a stronger coupling strength $\xi$ and thereby a larger decay
rate $\rm{Re}(\Upsilon)$ of the synthetic giant atom.

The giant-atom self-interference effect can instead be established
by plotting $\rho_{r_{1}r_{1},r_{2}r_{2}}(t)$ in
Fig.~\figpanel{fig2}{b} for $\Delta_{c}=30$ MHz and different values
of $\phi$. It is clear that an enhanced (reduced) decay occurs for
$\rho_{r_{1}r_{1},r_{2}r_{2}}(t)$ in the case of $\phi=2m\pi$
($\phi=2m\pi+\pi$) with $m\in\mathbb{Z}$ due to a perfect
constructive (destructive) interference between two coupling points,
as can be seen from $\rm{Im}(\Upsilon)=0$ and
$\rm{Re}(\Upsilon)=\Gamma(1+\cos{\phi})\Omega_{c}^{2}/\Delta_{c}^{2}$.
The atomic pair is found in particular to show an undamped
double-excitation population
[$\rho_{r_{1}r_{1},r_{2}r_{2}}(t)\equiv1$] for $\phi=2m\pi+\pi$ and
$\gamma=0$, which is one of the most remarkable features of giant
atoms~\cite{2014} due to a complete decoupling from the
waveguide ($\Upsilon=0$) and a vanishing intrinsic decay
($\gamma=0$). In the case of $\phi=2m\pi\pm\pi/2$, we have
$\rm{Im}(\Upsilon)=\pm\Gamma\Omega_{c}^{2}/\Delta_{c}^{2}$ and
$\rm{Re}(\Upsilon)=\Gamma\Omega_{c}^{2}/\Delta_{c}^{2}$, which
accounts for the identical population dynamics with a moderate decay
since opposite detunings (Lamb shifts) make no difference.

We finally detail how one can actually adjust phase $\phi$ while
leaving potential $V_{6}$ unchanged. To this end, a pair of
$^{87}$Rb atoms with ground state
$|g_{1,2}\rangle=|5S_{1/2},F=2,m_{F}=2\rangle$ and Rydberg state
$|r_{1,2}\rangle=|75P_{3/2},m_{J}=3/2\rangle$ of transition
frequency $\omega_{e}\simeq2\pi\times1009$ THz are taken here as an
example. In this case, we have $\gamma\simeq1.0$ kHz for the
intrinsic Rydberg lifetime $\tau\simeq964$ $\mu$s while
$V_{6}\simeq20$ GHz for $R\simeq3.1$ $\mu$m and
$C_{6}\simeq2\pi\times2.8\times10^{12}$
s$^{-1}\mu$m$^{6}$~\cite{prarydberg1,website1}. When the atomic pair
is placed exactly along the waveguide, we have $d=R$ and hence
$\phi\simeq41.6\pi$ by assuming that $v_{g}$ is a half of the vacuum
light speed $c$. When the atomic pair is misaligned along the
waveguide, however, it is viable to attain $d\simeq2.95$ $\mu$m and
hence $\phi\simeq39.6\pi$ with neither $R$ nor $V_{6}$
changed~\cite{supp}.

\begin{figure}[ptb]
\centering
\includegraphics[width=8.5 cm]{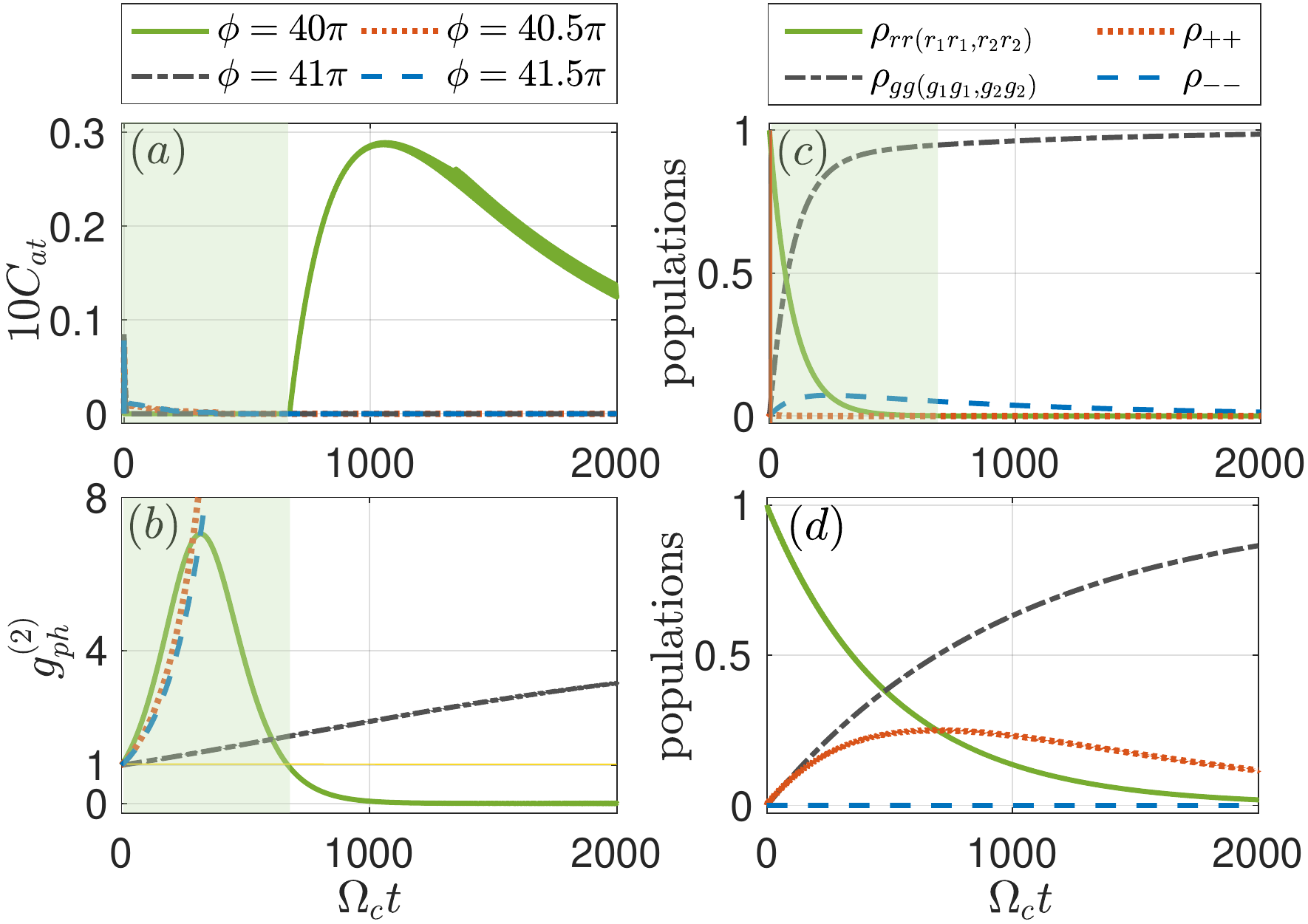}
\caption{Time evolutions of atomic concurrence $C_{at}$ (a) and
photonic correlation $g^{(2)}_{ph}$ for different values of $\phi$
(b) as well as atomic populations $\rho_{gg}$, $\rho_{rr}$,
$\rho_{++}$, and $\rho_{--}$ for $\phi=40\pi$ (c) and $\phi=41\pi$
(d) with other parameters same as in Fig.~\ref{fig2}(b).}
\label{fig3}
\end{figure}

\vspace{1mm}

\textit{Atomic entanglement onset: the long-time regime}.---Now
leaving the short-time regime where the atomic pair can be modeled
as a giant atom according to Eq.~(\ref{master2}), we turn to the
long-time regime where the decay toward non-guided vacuum modes
becomes important as less and less population can be found in the
double-excitation state. A peculiar aspect of the long-time regime
is internal entanglement of the atomic pair (i.e., a specific
superposition of two single-excitation states) generated by the
coherent interaction described by $H_{at}$, which can be quantified
by the Wootters concurrence~\cite{entanglement1,entanglement2,2023prl}
\begin{equation}
\begin{split}
C_{at}=Max(0,\lambda_{1}-\lambda_{2}-\lambda_{3}-\lambda_{4})
\label{Cat}
\end{split}
\end{equation}
with $\lambda_{1}>\lambda_{2}>\lambda_{3}>\lambda_{4}$ being the
four eigenvalues of matrix $X$ defined by
$X^{2}=\sqrt{\rho}(\sigma_{y}\otimes\sigma_{y})\rho^{*}(\sigma_{y}\otimes\sigma_{y})\sqrt{\rho}$
and the standard Pauli matrix $\sigma_{y}$. This concurrence takes
values in the range of $[0,1]$ with $C_{at}=0$ ($C_{at}=1$) denoting
a non-entangled (maximally entangled) state and is related to the
correlation function~\cite{2006pra,2010pra,2012prl}
\begin{equation}
\begin{split}
g^{(2)}_{ph}=\frac{\rho_{r_{1}r_{1},r_{2}r_{2}}}{\rho_{r_{1}r_{1}}\rho_{r_{2}r_{2}}}
\label{g2}
\end{split}
\end{equation}
of the photons emitted by two Rydberg atoms. As usual,
$\rho_{r_{1}r_{1}}=\langle r_{1}|\text{Tr}_{2}\rho|r_{1}\rangle$ and
$\rho_{r_{2}r_{2}}=\langle r_{2}|\text{Tr}_{1}\rho|r_{2}\rangle$ are
obtained from reduced density matrices of different atoms, while
$g^{(2)}_{ph}>1$ and $g^{(2)}_{ph}<1$ refer to the effects of photon
bunching and anti-bunching, respectively.

We plot in Fig.~\figpanel{fig3}{a} time evolutions of $C_{at}$ for
different values of $\phi$ starting from the same double-excitation
state and find that $C_{at}$ becomes suddenly nonzero at a critical
time for $\phi=2m\pi$ but remains vanishing for other values of
$\phi$. Such an onset of internal atomic entanglement happens when
the photon correlation function $g^{(2)}_{ph}$ evolves from the
regime of bunching to that of anti-bunching as shown in
Fig.~\figpanel{fig3}{b}. This signifies a rigid correspondence
between the emergence of photon anti-bunching and the generation of
atomic entanglement.

It is worth noting that $g^{(2)}_{ph}$ also indicates how the two
Rydberg atoms are distributed in the single- and double-excitation
states, which would be helpful to understand the underlying physics
of the entanglement sudden-onset dynamics. This inspires us to
further plot $\rho_{gg}$, $\rho_{rr}$, $\rho_{++}$, and $\rho_{--}$
in Fig.~\figpanel{fig3}{c} for $\phi=2m\pi$ and
Fig.~\figpanel{fig3}{d} for $\phi=2m\pi+\pi$ with
$|g\rangle\equiv|g_{1}g_{2}\rangle$,
$|r\rangle\equiv|r_{1}r_{2}\rangle$, and $|\pm
\rangle\equiv1/\sqrt2(|r_{1}g_{2} \rangle\pm |g_{1}r_{2} \rangle)$
by considering that $|+\rangle$ ($|-\rangle$) is the symmetric
(anti-symmetric) dressed state of $H_{at}$ in the case of
$\phi=m\pi$ due to $J_{ex}=0$. To be more specific, we have the
following dynamic equations~\cite{supp}
\begin{equation}
\begin{split}
\partial_{t}\rho_{++}&=-\gamma_{+}\rho_{++}+\gamma\rho_{rr}+i\Omega_{c}^{\ast}\rho_{r+}-i\Omega_{c}\rho_{+r},\\
\partial_{t}\rho_{--}&=-\gamma_{-}\rho_{--}+\gamma\rho_{rr},
\label{eq8}
\end{split}
\end{equation}
with $\gamma_{\pm}=\gamma+\Gamma\pm\Gamma_{ex}$. It is clear that
$|-\rangle$ is decoupled from field $\Omega_{c}$ and will become a
\emph{dark} state if it is further immune to the waveguide modes in
the case of $\phi=2m\pi$. However, $|+\rangle$ is always a
\emph{bright} state in that its dynamics depends on field
$\Omega_{c}$ all the time.

Note in particular that
$\gamma_{+}=\gamma+2\Gamma\gg\gamma_{-}=\gamma$ in the case of
$\phi=2m\pi$, which explains why $\rho_{++}$ remains to be vanishing
while $\rho_{--}$ does not in Fig.~\figpanel{fig3}{c} so that we
have $g^{(2)}_{ph}\simeq4\rho_{rr}/\rho_{--}^{2}$. The atomic
entanglement onset occurs for $\phi=2m\pi$ just because a dark state
immune to field $\Omega_{c}$ allows the transition from
$4\rho_{rr}>\rho_{--}^{2}$ to $4\rho_{rr}<\rho_{--}^{2}$. In the
case of $\phi=2m\pi+\pi$, however, we have a nonzero $\rho_{++}$ and
a vanishing $\rho_{--}$ in Fig.~\figpanel{fig3}{d} and hence
$g^{(2)}_{ph}\simeq4\rho_{rr}/\rho_{++}^{2}$ due to
$\gamma_{-}=\gamma+2\Gamma\gg\gamma_{+}=\gamma$. The atomic
entanglement onset is absent for $\phi=2m\pi+\pi$ just because a
bright state interacting with field $\Omega_{c}$ always results in
$4\rho_{rr}>\rho_{++}^{2}$. Finally, we stress that $\rho_{++}>0$
($\rho_{--}>0$) and $\rho_{--}=0$ ($\rho_{++}=0$) do not mean that
the decomposition of a mixed state $\rho$ must include a pure state
$|+\rangle$ ($|-\rangle$), hence do not mean that we must have
$C_{at}>0$, which further explains why $C_{at}$ could suddenly
become nonzero only for $\phi=2m\pi$.

\vspace{1mm}

\begin{figure}[ptb]
\centering
\includegraphics[width=8.5 cm]{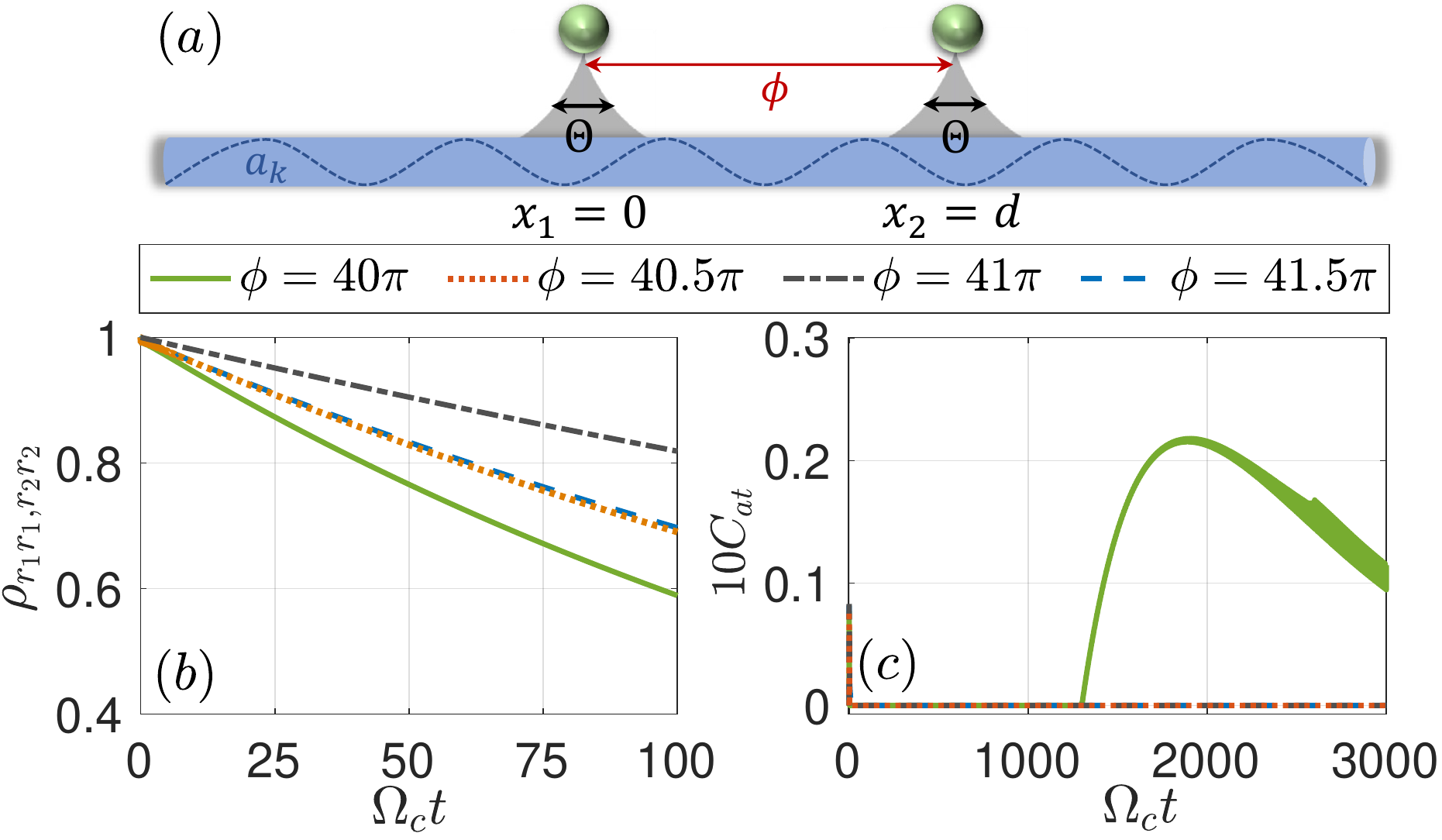}
\caption{(a) Schematic of two Rydberg atoms close to a waveguide
with identical exponential coupling distributions around $x_{1}=0$
and $x_{2}=d$. Time evolutions of population
$\rho_{r_{1}r_{1},r_{2}r_{2}}$ (b) and concurrence $C_{at}$ (c) with
$\Theta=5\pi/2$ and different values of $\phi$. Other parameters are
the same as in Fig.~\ref{fig2}(b).} \label{fig4}
\end{figure}

\textit{Continuous couplings}.---Working with point-like
atom-waveguide couplings, as assumed so far, is just a rough
approximation for highly-excited Rydberg states of size
$\bar{r}\propto n^{2}$, with $n$ being the principal quantum number.
We then extend our discrete-coupling configuration results to the
continuum limit whereby the coupling region becomes a large ensemble
of coupling points, each with a different strength. For an
exponential continuous distribution~\cite{continuum} of such
coupling strengths as in Fig.~\figpanel{fig4}{a} spread about each
contact point with a characteristic width $\Theta$, we find that the
master equations~(\ref{master4}) and (\ref{master2}) can be
generalized to the continuous-coupling limit by replacing $\Gamma
\to \Gamma^{\prime}$, $\Gamma_{ex}\to \Gamma_{ex}^{\prime}$,
$J_{ex}\to J_{ex}^{\prime}$ and $\Upsilon\to \Upsilon'$ while
simultaneously introducing an interaction term
$J'(\sigma_{+}^{1}\sigma_{-}^{1}+\sigma_{+}^{2}\sigma_{-}^{2})$ in
Eq.~(\ref{Hat}). While explicit expressions of these modified
parameters are given in~\cite{supp}, we here use them to plot in
Figs.~\figpanel{fig4}{b} and \figpanel{fig4}{c} the time evolutions
of $\rho_{r_{1}r_{1},r_{2}r_{2}}$ and $C_{at}$ in the short-time and
long-time regimes, respectively, for different values of $\phi$ and
a fixed $\Theta$.

It is easy to see that the giant-atom effects of phase-dependent
population decay and entanglement onset remain observable for a
remarkable coupling broadening. Moreover, the dynamic behavior of
$\rho_{r_{1}r_{1},r_{2}r_{2}}$ for $\Theta=5\pi/2$ (continuous
couplings) is identical to that for $\Theta=0$ (point-like
couplings) in the case of $\phi=2m\pi+\pi$ since there is no decay
toward the waveguide with $\rm{Re}(\Upsilon')=0$. Note, however,
that a nonzero Lamb shift with $\rm{Im}(\Upsilon')\ne0$ always
exists for continuous couplings~\cite{supp}, which does not affect
atomic population decay but would be relevant to other problems such
as photon scattering.

\vspace{1mm}

\textit{Conclusions}.---We have proposed a feasible scheme for
constructing a synthetic giant atom with two interacting Rydberg
atoms coupled to a PCW and driven by a coherent field. Giant-atom
effects are manifested by a phase-dependent population dynamics in
the short-time regime and an entanglement sudden-onset dynamics in
the long-time regime. These effects are observable even for
continuous atom-waveguide couplings, indicating their robustness
against the unavoidable coupling broadening. Compared to typical
schemes utilizing superconducting quantum circuits, our Rydberg
scheme provides a promising platform for studying giant-atom physics
in the optical regime. Hence, our findings have potential
applications in quantum network engineering and quantum information
processing based on optical photons.

\acknowledgments{This work is supported by the National Natural
Science Foundation of China (No. 12074061), the National Key
Research and Development Program of China (No. 2021YFE0193500), and
the Italian PNRR MUR (project PE0000023-NQSTI).}

\clearpage

\onecolumngrid

\begin{center}
{\large \textbf{Supplemental Material for ``Giant-Atom Effects on Population and Entanglement Dynamics of Rydberg Atoms''}}

\vspace{8mm}

Yao-Tong Chen$^{1}$, Lei Du$^{1,\blue{*}}$, Yan Zhang$^{1}$, Lingzhen Guo$^{2}$,  Jin-Hui Wu$^{1,\blue{\dagger}}$, M. Artoni$^{3,4}$, and G. C. La Rocca$^{5}$

\end{center}

\begin{minipage}[]{16cm}
\small{\it

\centering $^{1}$ School of Physics and Center for Quantum Sciences,
Northeast Normal University, Changchun 130024, China  \\
\centering $^{2}$ Center for Joint Quantum Studies and Department of
Physics, School of Science, Tianjin University, Tianjin 300072,
China  \\
\centering $^{3}$ Department of Chemistry and Physics of Materials,
University of Brescia, Italy \\
\centering $^{4}$ European Laboratory for
Non-Linear Spectroscopy, Sesto Fiorentino, Italy \\
\centering $^{5}$ NEST, Scuola Normale Superiore, Piazza dei Cavalieri 7,
I-56126 Pisa, Italy \\
}

\end{minipage}

\vspace{8mm}

This supplementary material gives further details on the misaligned
arrangement of two atoms along a waveguide (Sec. I), the master
equation of a two-atom four-level configuration (Sec. II), the
master equation of a giant-atom two-level configuration (Sec. III),
and the continuous couplings of Rydberg atoms and waveguide modes
(Sec. IV) that are omitted in the main text.

\section{I.\quad Misaligned arrangement of two atoms along a waveguide}

\begin{figure}[pth]
\centering
\includegraphics[width=10 cm]{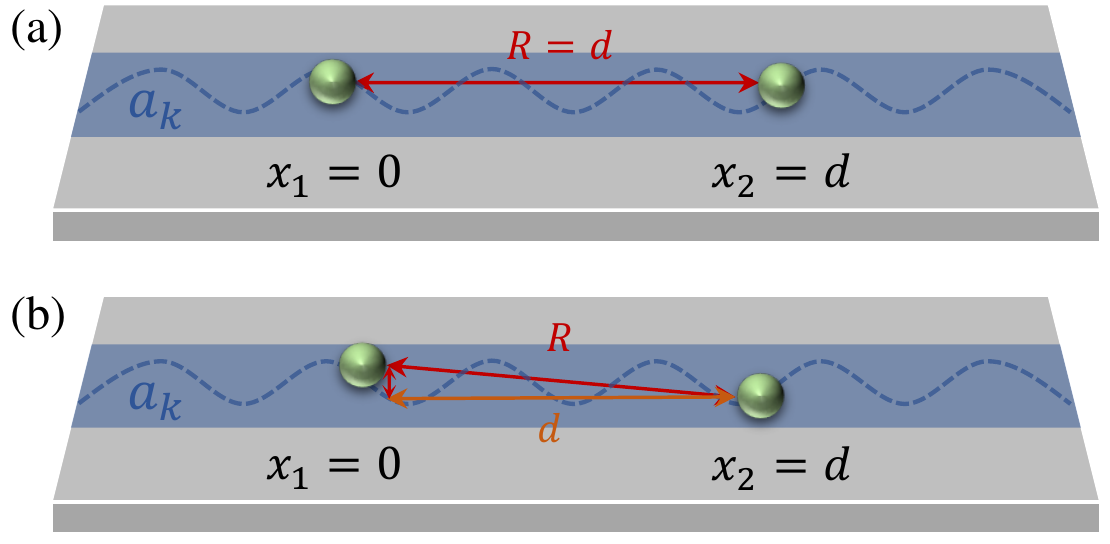}
\renewcommand\thefigure{S1}
\caption{Arrangement details of two Rydberg atoms with respect to a
one-dimensional waveguide shown as a blue area. The atomic pair is
(a) placed along the waveguide with $d=R$; (b) misaligned along the
waveguide with $d<R$. }\label{FIGS1}
\end{figure}

In this section, we discuss how to manipulate the van der Waals
(vdW) potential $V_{6}$ and the accumulated phase $\phi$ separately.
As mentioned in the main text, $V_{6}$ and $\phi$ are determined by
the (straight-line) distance $R$ between a pair of Rydberg atoms and
the separation $d$ (along the waveguide) between two coupling
points, respectively. When this atomic pair is placed exactly along
the waveguide, we have $R\equiv d$ so that $\phi$ or $V_{6}$ cannot
be changed alone as shown in Fig.~\figpanel{FIGS1}{a}. In this case,
it is easy to attain
$\phi\simeq\omega_{e}d/v_{g}=\omega_{e}R/v_{g}=41.6\pi$ with
$R\simeq3.1\,\mu\text{m}$,
$\omega_{e}\simeq2\pi\times1009\,\text{THz}$, and $v_{g}\simeq0.5c$
as considered in the main text. In order to change $\phi$ and
$V_{6}$ separately, we can choose a \emph{misaligned} arrangement of
this atomic pair as shown in Fig.~\figpanel{FIGS1}{b}, where $d$ is
clearly smaller than $R$. In this case, keeping
$R\simeq3.1\,\mu\text{m}$ and hence $V_{6}=20\,\text{GHz}$
unchanged, it is viable to tune $\phi$ in the range of $[41.6,
39.6]\pi$ by reducing $d$ from $3.1\,\mu\text{m}$ with a vanishing
misaligned angle to $2.95\,\mu\text{m}$ with a $0.09\pi$ misaligned
angle.

\section{II.\quad master equation of a two-atom four-level configuration}

In this section, we provide the derivation procedures from Eq.~(1)
on Hamiltonian $H$ to Eq.~(2) on density operator $\rho$ in the main
text. As shown in Fig~1(b), two upper transitions
$|g_{1}r_{2}\rangle\leftrightarrow|r_{1}r_{2} \rangle$ and
$|r_{1}g_{2}\rangle\leftrightarrow|r_{1}r_{2} \rangle$ are driven by
the external coherent field $\Omega_{c}$ while two lower transitions
$|g_{1}g_{2}\rangle\leftrightarrow|g_{1}r_{2} \rangle$ and
$|g_{1}g_{2}\rangle\leftrightarrow|r_{1}g_{2} \rangle$ are coupled
to the waveguide modes. Under the two-photon resonance condition
(i.e., $\Delta_{c}\simeq-\delta_{k}$ with
$\Delta_{c}=\omega_{c}-\omega_{e}-V_{6}$ and
$\delta_{k}=\omega_{k}-\omega_{e}$), we neglect the interactions
resulted from coherent field $\omega_{c}$ in $H$ for the moment and
move to the interaction picture with respect to
$H_{0}=(2\omega_{e}+V_{6}-\omega_{c})(\sigma_{+}^{1}\sigma_{-}^{1}+\sigma_{+}^{2}\sigma_{-}^{2})+\int
dk\omega_{k}a_{k}^{\dagger}a_{k}=
(\omega_{e}-\Delta_{c})(\sigma_{+}^{1}\sigma_{-}^{1}+\sigma_{+}^{2}\sigma_{-}^{2})+\int
dk\omega_{k}a_{k}^{\dagger}a_{k}$. Then the Hamiltonian describing
the atom-waveguide interaction can be written as
\begin{equation}
\begin{split}
H_{\text{int}}(t)&=\int_{-\infty}^{+\infty}dk\left[ge^{-i(\Delta_{c}+\omega_{k}-\omega_{e})t}a_{k}\sigma_{+}^{1}
+ge^{ikd}e^{-i(\Delta_{c}+\omega_{k}-\omega_{e})t}a_{k}\sigma_{+}^{2}+\text{H.c.}\right]\\
&=\int_{-\infty}^{+\infty}dk\left[ge^{-i(\Delta_{c}+\delta_{k})t}a_{k}\sigma_{+}^{1}
+ge^{ikd}e^{-i(\Delta_{c}+\delta_{k})t}a_{k}\sigma_{+}^{2}+\text{H.c.}\right]
\end{split}
\tag{S1}
\label{intH}
\end{equation}
with $\sigma_{+}^{1}=(\sigma_{-}^{1})^{\dagger}=|r_{1}g_{2}
\rangle\langle g_{1}g_{2}|$ and
$\sigma_{+}^{2}=(\sigma_{-}^{2})^{\dagger}=|g_{1}r_{2}
\rangle\langle g_{1}g_{2}|$ defined as in the main text.

To study the population dynamics of this atomic pair, we can
eliminate the waveguide field via a standard procedure and calculate
the following master equation for a reduced density
operator~\cite{book,2016}
\begin{equation}
\begin{split}
\partial_{t}\rho(t)&=-\int_{0}^{\infty}d\tau \text{Tr}_{w}{[H_{\text{int}}(t),[H_{\text{int}}(t-\tau),\rho_{w}\otimes\rho(t)]]},\\
\end{split}
\tag{S2}
\label{rho}
\end{equation}
where $\text{Tr}_{w}$ represents a partial tracing over the
waveguide degrees of freedom and $\rho_{w}=|0 \rangle\langle 0|$ is
the \textit{initial} vacuum state of the waveguide modes.
Substituting Eq.~(\ref{intH}) into Eq.~(\ref{rho}) we have
\begin{equation}
\begin{split}
\partial_{t}\rho&=\sum_{j=1,2}\Gamma\left(\sigma_{-}^{j}\rho\sigma_{+}^{j}-\frac{1}{2}\sigma_{+}^{j}\sigma_{-}^{j}\rho-\frac{1}{2}\rho\sigma_{+}^{j}\sigma_{-}^{j}\right)
+\frac{\tilde{\Gamma}+\tilde{\Gamma}^{*}}{2}\left(\sigma_{-}^{1}\rho\sigma_{+}^{2}+\sigma_{-}^{2}\rho\sigma_{+}^{1}\right)\\
&\quad\,-\frac{\tilde{\Gamma}}{2}\left(\sigma_{+}^{1}\sigma_{-}^{2}\rho+\sigma_{+}^{2}\sigma_{-}^{1}\rho\right)-\frac{\tilde{\Gamma}^{*}}{2}\left(\rho\sigma_{+}^{1}\sigma_{-}^{2}+\rho\sigma_{+}^{2}\sigma_{-}^{1}\right)\\
&=\sum_{j=1,2}\Gamma\left(\sigma_{-}^{j}\rho\sigma_{+}^{j}-\frac{1}{2}\sigma_{+}^{j}\sigma_{-}^{j}\rho-\frac{1}{2}\rho\sigma_{+}^{j}\sigma_{-}^{j}\right)
+\Gamma_{ex}\left(\sigma_{-}^{1}\rho\sigma_{+}^{2}+\sigma_{-}^{2}\rho\sigma_{+}^{1}\right)\\
&\quad\,-\frac{\Gamma_{ex}}{2}\left(\sigma_{+}^{1}\sigma_{-}^{2}\rho+\sigma_{+}^{2}\sigma_{-}^{1}\rho+\rho\sigma_{+}^{1}\sigma_{-}^{2}+\rho\sigma_{+}^{2}\sigma_{-}^{1}\right)
-\frac{iJ_{ex}}{2}\left(\sigma_{+}^{1}\sigma_{-}^{2}\rho+\sigma_{+}^{2}\sigma_{-}^{1}\rho-\rho\sigma_{+}^{1}\sigma_{-}^{2}-\rho\sigma_{+}^{2}\sigma_{-}^{1}\right),
\end{split}
\tag{S3}
\label{rho1}
\end{equation}
with
\begin{equation}
\begin{split}
\Gamma&=2g^{2}\int_{0}^{\infty}d\tau \int_{-\infty}^{+\infty}dke^{\pm i(\Delta_{c}+\delta_{k})\tau}
=2g^{2}\int_{0}^{\infty}d\tau e^{\pm i\Delta_{c}\tau}\left(\int_{-\infty}^{0}dke^{\pm i\delta_{k}\tau}+\int_{0}^{+\infty}dke^{\pm i\delta_{k}\tau}\right)\\
&=\frac{4g^{2}}{v_{g}}\int_{0}^{\infty}d\tau e^{\pm i\Delta_{c}\tau}\int_{-\infty}^{+\infty}d\delta_{k}e^{\pm i\delta_{k}\tau}=\frac{4g^{2}}{v_{g}}\int_{0}^{\infty}d\tau e^{\pm i\Delta_{c}\tau}2\pi\delta(\tau)=\frac{4\pi g^{2}}{v_{g}}=4\pi g^{2}D(k),\\
\tilde{\Gamma}&=2g^{2}\int_{0}^{\infty}d\tau
\int_{-\infty}^{+\infty}dke^{-i(\Delta_{c}+\delta_{k})\tau}e^{\pm
ikd}\\
&=2g^{2}\int_{0}^{\infty}d\tau\left[\int_{-\infty}^{0}dke^{-i(\Delta_{c}+\delta_{k})\tau}e^{\mp i(\delta_{k}+\omega_{e})d/v_{g}}+\int_{0}^{+\infty}dke^{-i(\Delta_{c}+\delta_{k})\tau}e^{\pm i(\delta_{k}+\omega_{e})d/v_{g}}\right]\\
&=\frac{2g^{2}}{v_{g}}\int_{0}^{\infty}d\tau\left[\int_{-\infty}^{+\infty}d\delta_{k}e^{-i(\Delta_{c}+\delta_{k})\tau}e^{\mp i(\delta_{k}+\omega_{e})d/v_{g}}+\int_{-\infty}^{+\infty}d\delta_{k}e^{-i(\Delta_{c}+\delta_{k})\tau}e^{\pm i(\delta_{k}+\omega_{e})d/v_{g}}\right]\\
&=\frac{2g^{2}}{v_{g}}\left[e^{i(\omega_{e}d/v_{g}-\Delta_{c}\tau)}\int_{0}^{\infty}d\tau 2\pi\delta\left(\tau-d/v_{g}\right)+e^{-i(\omega_{e}d/v_{g}+\Delta_{c}\tau)}\int_{0}^{\infty}d\tau 2\pi\delta\left(\tau+d/v_{g}\right)\right]\\
&=\frac{4\pi
g^{2}}{v_{g}}e^{i(\omega_{e}-\Delta_{c})d/v_{g}}\simeq\frac{4\pi
g^{2}}{v_{g}}e^{i\omega_{e}d/v_{g}}=\Gamma e^{i\phi},
\end{split}
\tag{S4}
\label{gam}
\end{equation}
as well as
$\Gamma_{ex}=\text{Re}\{\tilde{\Gamma}\}=\Gamma\text{cos}\phi$ and
$J_{ex}=\text{Im}\{\tilde{\Gamma}\}=\Gamma\text{sin}\phi$. In the
above derivation, we have also considered the $\delta$ function
definition $\int_{-\infty}^{+\infty}dke^{\pm ikx}=2\pi\delta(x)$ and
the waveguide mode density $D(k)=\partial
k/\partial\omega_{k}$~\cite{2014Kockum,2016Longhi,nature2020}.

Note that Eq.~(\ref{rho1}) just describes the interactions between a
continuum of waveguide modes and two lower atomic transitions
$|g_{1}g_{2}\rangle\leftrightarrow|g_{1}r_{2} \rangle$ and
$|g_{1}g_{2}\rangle\leftrightarrow|r_{1}g_{2} \rangle$. Further
taking into account the intrinsic atomic decay toward non-guided
modes in the free space as well as the neglected interactions
between a coherent field and two upper atomic transitions
$|g_{1}r_{2}\rangle\leftrightarrow|r_{1}r_{2} \rangle$ and
$|r_{1}g_{2}\rangle\leftrightarrow|r_{1}r_{2} \rangle$ in $H$, one
can easily obtain the master equation (2) in the main text. This
equation, if expanded in the two-atom four-level configuration, will
turn out to be
\begin{equation}
\begin{split}
\partial_{t}\rho_{g_{1}g_{1},g_{2}g_{2}}&=(\gamma+\Gamma)\rho_{r_{1}r_{1},g_{2}g_{2}}+(\gamma+\Gamma)\rho_{g_{1}g_{1},r_{2}r_{2}}
+\Gamma_{ex}\rho_{r_{1}g_{1},g_{2}r_{2}}+\Gamma_{ex}\rho_{g_{1}r_{1},r_{2}g_{2}},\\
\partial_{t}\rho_{r_{1}r_{1},g_{2}g_{2}}&=-(\gamma+\Gamma)\rho_{r_{1}r_{1},g_{2}g_{2}}+\gamma\rho_{r_{1}r_{1},r_{2}r_{2}}-\tilde{\Gamma}/2\rho_{g_{1}r_{1},r_{2}g_{2}}-\tilde{\Gamma}^{*}/2\rho_{r_{1}g_{1},g_{2}r_{2}}
-i(\Omega_{c}\rho_{r_{1}r_{1},g_{2}r_{2}}-\Omega_{c}^{*}\rho_{r_{1}r_{1},r_{2}g_{2}}),\\
\partial_{t}\rho_{g_{1}g_{1},r_{2}r_{2}}&=-(\gamma+\Gamma)\rho_{g_{1}g_{1},r_{2}r_{2}}
+\gamma\rho_{r_{1}r_{1},r_{2}r_{2}}-\tilde{\Gamma}/2\rho_{r_{1}g_{1},g_{2}r_{2}}-\tilde{\Gamma}^{*}/2\rho_{g_{1}r_{1},r_{2}g_{2}}
-i(\Omega_{c}\rho_{g_{1}r_{1},r_{2}r_{2}}-\Omega_{c}^{*}\rho_{r_{1}g_{1},r_{2}r_{2}}),\\
\partial_{t}\rho_{g_{1}r_{1},g_{2}g_{2}}&=(i\Delta_{c}-\gamma/2-\Gamma/2)\rho_{g_{1}r_{1},g_{2}g_{2}}+\gamma\rho_{g_{1}r_{1},r_{2}r_{2}}-\tilde{\Gamma}^{*}/2\rho_{g_{1}g_{1},g_{2}r_{2}}
-i\Omega_{c}\rho_{g_{1}r_{1},g_{2}r_{2}},\\
\partial_{t}\rho_{r_{1}g_{1},g_{2}g_{2}}&=-(i\Delta_{c}+\gamma/2+\Gamma/2)\rho_{r_{1}g_{1},g_{2}g_{2}}+\gamma\rho_{r_{1}g_{1},r_{2}r_{2}}-\tilde{\Gamma}/2\rho_{g_{1}g_{1},r_{2}g_{2}}+i\Omega_{c}^{*}\rho_{r_{1}g_{1},r_{2}g_{2}},\\
\partial_{t}\rho_{g_{1}g_{1},g_{2}r_{2}}&=(i\Delta_{c}-\gamma/2-\Gamma/2)\rho_{g_{1}g_{1},g_{2}r_{2}}+\gamma\rho_{r_{1}r_{1},g_{2}r_{2}}-\tilde{\Gamma}^{*}/2\rho_{g_{1}r_{1},g_{2}g_{2}}
-i\Omega_{c}\rho_{g_{1}r_{1},g_{2}r_{2}},\\
\partial_{t}\rho_{g_{1}g_{1},r_{2}g_{2}}&=-(i\Delta_{c}+\gamma/2+\Gamma/2)\rho_{g_{1}g_{1},r_{2}g_{2}}+\gamma\rho_{r_{1}r_{1},r_{2}g_{2}}-\tilde{\Gamma}/2\rho_{r_{1}g_{1},g_{2}g_{2}}
+i\Omega_{c}^{*}\rho_{r_{1}g_{1},r_{2}g_{2}},\\
\partial_{t}\rho_{g_{1}r_{1},g_{2}r_{2}}&=-\gamma\rho_{g_{1}r_{1},g_{2}r_{2}}
-i\Omega_{c}^{*}(\rho_{g_{1}r_{1},g_{2}g_{2}}+\rho_{g_{1}g_{1},g_{2}r_{2}}),\\
\partial_{t}\rho_{r_{1}g_{1},r_{2}g_{2}}&=-\gamma\rho_{r_{1}g_{1},r_{2}g_{2}}
+i\Omega_{c}(\rho_{r_{1}g_{1},g_{2}g_{2}}+\rho_{g_{1}g_{1},r_{2}g_{2}}),\\
\partial_{t}\rho_{r_{1}g_{1},g_{2}r_{2}}&=-(\gamma+\Gamma)\rho_{r_{1}g_{1},g_{2}r_{2}}-\tilde{\Gamma}/2\rho_{g_{1}g_{1},r_{2}r_{2}}-\tilde{\Gamma}^{*}/2\rho_{r_{1}r_{1},g_{2}g_{2}}
-i(\Omega_{c}\rho_{r_{1}r_{1},g_{2}r_{2}}-\Omega_{c}^{*}\rho_{r_{1}g_{1},r_{2}r_{2}}),\\
\partial_{t}\rho_{g_{1}r_{1},r_{2}g_{2}}&=-(\gamma+\Gamma)\rho_{g_{1}r_{1},r_{2}g_{2}}-\tilde{\Gamma}^{*}/2\rho_{g_{1}g_{1},r_{2}r_{2}}-\tilde{\Gamma}/2\rho_{r_{1}r_{1},g_{2}g_{2}}
+i(\Omega_{c}^{*}\rho_{r_{1}r_{1},r_{2}g_{2}}-\Omega_{c}\rho_{g_{1}r_{1},r_{2}r_{2}}),\\
\partial_{t}\rho_{r_{1}r_{1},g_{2}r_{2}}&=-(i\Delta_{c}+3\gamma/2+\Gamma/2)\rho_{r_{1}r_{1},g_{2}r_{2}}-\tilde{\Gamma}/2\rho_{g_{1}r_{1},r_{2}r_{2}}
-i\Omega_{c}^{*}(\rho_{r_{1}r_{1},g_{2}g_{2}}+\rho_{r_{1}g_{1},g_{2}r_{2}}-\rho_{r_{1}r_{1},r_{2}r_{2}}),\\
\partial_{t}\rho_{r_{1}r_{1},r_{2}g_{2}}&=(i\Delta_{c}-3\gamma/2-\Gamma/2)\rho_{r_{1}r_{1},r_{2}g_{2}}-\tilde{\Gamma}^{*}/2\rho_{r_{1}g_{1},r_{2}r_{2}}
+i\Omega_{c}(\rho_{r_{1}r_{1},g_{2}g_{2}}+\rho_{g_{1}r_{1},r_{2}g_{2}}-\rho_{r_{1}r_{1},r_{2}r_{2}}),\\
\partial_{t}\rho_{g_{1}r_{1},r_{2}r_{2}}&=-(i\Delta_{c}+3\gamma/2+\Gamma/2)\rho_{g_{1}r_{1},r_{2}r_{2}}-\tilde{\Gamma}/2\rho_{r_{1}r_{1},g_{2}r_{2}}
-i\Omega_{c}^{*}(\rho_{g_{1}r_{1},r_{2}g_{2}}+\rho_{g_{1}g_{1},r_{2}r_{2}}-\rho_{r_{1}r_{1},r_{2}r_{2}}),\\
\partial_{t}\rho_{r_{1}g_{1},r_{2}r_{2}}&=(i\Delta_{c}-3\gamma/2-\Gamma/2)\rho_{r_{1}g_{1},r_{2}r_{2}}-\tilde{\Gamma}^{*}/2\rho_{r_{1}r_{1},r_{2}g_{2}}
+i\Omega_{c}(\rho_{r_{1}g_{1},g_{2}r_{2}}+\rho_{g_{1}g_{1},r_{2}r_{2}}-\rho_{r_{1}r_{1},r_{2}r_{2}}),
\end{split}
\tag{S5}
\label{mas}
\end{equation}
constrained by
$\rho_{g_{1}g_{1},g_{2}g_{2}}+\rho_{r_{1}r_{1},g_{2}g_{2}}+\rho_{g_{1}g_{1},r_{2}r_{2}}+\rho_{r_{1}r_{1},r_{2}r_{2}}=1$.

As mentioned in the main text, it is helpful to understand the
long-time entanglement onset dynamics by replacing single-excitation
states $|g_{1}r_{2}\rangle$ and $|r_{1}g_{2}\rangle$ with their
superpositions $|\pm\rangle=1/\sqrt2(|r_{1}g_{2} \rangle\pm
|g_{1}r_{2} \rangle)$. Population evolutions in the two symmetric
and anti-symmetric states can be calculated from the above equations
as
\begin{equation}
\begin{split}
\partial_{t}\rho_{++}&=\frac{1}{2}(\partial_{t}\rho_{g_{1}g_{1},r_{2}r_{2}}+\partial_{t}\rho_{r_{1}r_{1},g_{2}g_{2}}+\partial_{t}\rho_{r_{1}g_{1},g_{2}r_{2}}+\partial_{t}\rho_{g_{1}r_{1},r_{2}g_{2}})\\
&=\frac{1}{2}\left[(-\gamma-\Gamma-\tilde{\Gamma}^{*}/2-\tilde{\Gamma}/2)(\rho_{g_{1}g_{1},r_{2}r_{2}}+\rho_{r_{1}r_{1},g_{2}g_{2}}+\rho_{r_{1}g_{1},g_{2}r_{2}}+\rho_{g_{1}r_{1},r_{2}g_{2}})+
\gamma\rho_{r_{1}r_{1},r_{2}r_{2}}\right]\\
&\quad\,\,+i(\Omega_{c}^{*}\rho_{r_{1}r_{1},r_{2}g_{2}}+\Omega_{c}^{*}\rho_{r_{1}g_{1},r_{2}r_{2}}-\Omega_{c}\rho_{r_{1}r_{1},g_{2}r_{2}}-\Omega_{c}\rho_{g_{1}r_{1},r_{2}r_{2}})\\
&=-(\gamma+\Gamma+\Gamma_{ex})\rho_{++}+\gamma\rho_{rr}+i\Omega_{c}^{*}\rho_{r+}-i\Omega_{c}\rho_{+r},\\
\partial_{t}\rho_{--}&=\frac{1}{2}(\partial_{t}\rho_{g_{1}g_{1},r_{2}r_{2}}+\partial_{t}\rho_{r_{1}r_{1},g_{2}g_{2}}-\partial_{t}\rho_{r_{1}g_{1},g_{2}r_{2}}-\partial_{t}\rho_{g_{1}r_{1},r_{2}g_{2}})\\
&=\frac{1}{2}\left[(-\gamma-\Gamma+\tilde{\Gamma}^{*}/2+\tilde{\Gamma}/2)(\rho_{g_{1}g_{1},r_{2}r_{2}}+\rho_{r_{1}r_{1},g_{2}g_{2}}-\rho_{r_{1}g_{1},g_{2}r_{2}}-\rho_{g_{1}r_{1},r_{2}g_{2}})+
\gamma\rho_{r_{1}r_{1},r_{2}r_{2}}\right]\\
&=-(\gamma+\Gamma-\Gamma_{ex})\rho_{--}+\gamma\rho_{rr},\\
\end{split}
\tag{S6}
\label{+-}
\end{equation}
which are exactly Eq.~(8) in the main text if we further introduce
$\gamma_{\pm}=\gamma+\Gamma\pm\Gamma_{ex}$.

\section{III.\quad master equation of a giant-atom two-level configuration}

In the case that the single-excitation states $|r_{1}g_{2}\rangle$
and $|g_{1}r_{2}\rangle$ are not populated initially, if we have
$|\Delta_{c}|\gg\Omega_{c},g$ and $\Delta_{c}+\delta_{k}\simeq0$,
our two-atom four-level configuration can be reduced to a
(synthetic) giant-atom two-level configuration by eliminating
$|r_{1}g_{2}\rangle$ and $|g_{1}r_{2}\rangle$ in the
\textit{short-time} regime. In view of this, a pair of Rydberg atoms
will decay from the double-excitation state $|r_{1}r_{2}\rangle$
directly to the ground state $|g_{1}g_{2}\rangle$ by simultaneously
emitting a coherent-field photon of frequency $\omega_{c}$ and a
waveguide-mode photon of frequency $\omega_{k}$, through two
competing two-photon resonant transitions exhibiting effective
coupling strengths $\xi_1=-g\Omega_{c}/\Delta_{c}\equiv\xi$ and
$\xi_2=\xi e^{i\phi}$, respectively. This can be substantiated by
the following discussions starting from an \emph{effective}
Hamiltonian defined as~\cite{eff1,eff2}
\begin{equation}
\begin{split}
H_{e}(t)&=-iH_{I}(t)\int_{0}^{t} H_{I}(t')dt',
\end{split}
\tag{S7}
\label{H_{e}}
\end{equation}
with
\begin{equation}
\begin{split}
H_{I}(t)&=\int
dkga_{k}e^{-i(\Delta_{c}+\delta_{k})t}\left(\sigma_{+}^{1}+e^{ikd}\sigma_{+}^{2}\right)
+\Omega_{c}\left(\sigma_{+}^{3}+\sigma_{+}^{4}\right)+\text{H.c.}
\end{split}
\tag{S8}
\label{H_{I}}
\end{equation}
being the total interaction Hamiltonian involving both waveguide
modes and coherent field of our two-atom four-level configuration.
Substituting Eq.~(\ref{H_{I}}) into Eq.~(\ref{H_{e}}), one has
\begin{equation}
\begin{split}
H_{e}(t)&\simeq\frac{g^{2}}{\delta_{k}}\int dka_{k}a_{k}^{\dagger}(\sigma_{-}^{1}\sigma_{+}^{1}+\sigma_{-}^{2}\sigma_{+}^{2})-\frac{\Omega_{c}^{2}}{\Delta_{c}}(\sigma_{+}^{3}\sigma_{-}^{3}+\sigma_{+}^{4}\sigma_{-}^{4})\\
&\quad\,+\frac{g\Omega_{c}}{\delta_{k}}\int dka_{k}e^{-i(\delta_{k}+\Delta_{c})t}(\sigma_{+}^{3}\sigma_{+}^{1}+e^{ikd}\sigma_{+}^{4}\sigma_{+}^{2})-\frac{g\Omega_{c}}{\Delta_{c}}\int dka_{k}^{\dagger}e^{i(\delta_{k}+\Delta_{c})t}(\sigma_{-}^{1}\sigma_{-}^{3}+e^{-ikd}\sigma_{-}^{2}\sigma_{-}^{4})+\cdots\\
&=\frac{2g^{2}}{\delta_{k}}\int dka_{k}a_{k}^{\dagger}|g_{1}g_{2}\rangle \langle g_{1}g_{2}|-\frac{2\Omega_{c}^{2}}{\Delta_{c}}|r_{1}r_{2}\rangle \langle r_{1}r_{2}|\\
&\quad\,+\frac{g\Omega_{c}}{\delta_{k}}\int
dka_{k}e^{-i(\delta_{k}+\Delta_{c})t}(1+e^{ikd})|r_{1}r_{2}\rangle
\langle g_{1}g_{2}|-\frac{g\Omega_{c}}{\Delta_{c}}\int
dka_{k}^{\dagger}e^{i(\delta_{k}+\Delta_{c})t}(1+e^{-ikd})|g_{1}g_{2}\rangle
\langle r_{1}r_{2}|+\cdots,
\end{split}
\tag{S9}
\label{H_{ee}}
\end{equation}
where we have omitted a few terms related to the single-excited
states $|r_{1}g_{2}\rangle$ and $|g_{1}r_{2}\rangle$ since they are
decoupled from other states and only interact with each other.

Considering again $\Delta_{c}+\delta_{k}\simeq0$ and
$|\Delta_{c}|\gg\Omega_{c},g$ as mentioned above, we have
$|g^{2}/\delta_{k}|\to 0$ and $|\Omega_{c}^{2}/\Delta_{c}|\to 0$,
which then result in a reduction of $H_{e}$ into the interaction
Hamiltonian
\begin{equation}
\begin{split}
\mathcal{H}_{\text{int}}(t)=&\int_{-\infty}^{+\infty}dk\left[\xi(1+e^{ikd})e^{-i(\delta_{k}+\Delta_{c})t}a_{k}\sigma_{+}+\text{H.c.}\right],
\end{split}
\tag{S10} \label{intGA}
\end{equation}
for a synthetic giant atom with two levels
$|g\rangle=|g_{1}g_{2}\rangle$ and $|r\rangle=|r_{1}r_{2}\rangle$ by
taking $\xi=g\Omega_{c}/\delta_{k}=-g\Omega_{c}/\Delta_{c}$. We can
attain $\mathcal{H}$ in Eq.~(4) of the main text by rotating this
interaction Hamiltonian with respect to frequency
$2\omega_{e}+V_{6}$ of the giant-atom transition
$|g\rangle\leftrightarrow|r\rangle$, which is coupled to a continuum
of waveguide modes of frequency $\omega_{k}$ accompanied by a
coherent field of frequency $\omega_{c}$. Substituting
Eq.~(\ref{intGA}) into an equation similar to Eq.~(\ref{rho}) for
the giant-atom density operator $\varrho$, we can further attain the
master equation (5) in the main text by including also the intrinsic
atomic decay toward non-guided modes in the free space. This master
equation turns out to be
\begin{equation}
\begin{split}
\partial_{t}\varrho_{gg}&=(\Upsilon+\Upsilon^{*}+2\gamma)\varrho_{rr},\\
\partial_{t}\varrho_{gr}&=-(\Upsilon^{*}+\gamma)\varrho_{gr},\\
\partial_{t}\varrho_{rg}&=-(\Upsilon+\gamma)\varrho_{rg},
\end{split}
\tag{S11}
\label{intpp}
\end{equation}
after an expansion in the giant-atom two-level configuration and are
constrained by $\varrho_{gg}+\varrho_{rr}=1$ with
\begin{equation}
\begin{split}
\Upsilon&=2\xi^{2}\int_{0}^{\infty}d\tau \int_{-\infty}^{+\infty}dk\left[e^{\pm i(\delta_{k}+\Delta_{c})\tau}+e^{-i(\delta_{k}+\Delta_{c})\tau}e^{\pm ikd}\right]\\
&=4\pi\xi^{2}D(k)[1+e^{i\phi}]=(\Gamma+\Gamma_{ex}+iJ_{ex})\Omega_{c}^{2}/\Delta_{c}^{2}.
\end{split}
\tag{S12}
\label{gam}
\end{equation}

\section{IV.\quad continuous couplings of Rydberg atoms and waveguide modes}
In this section, we try to derive the explicit expressions of
relevant constants describing various interactions between two
Rydberg atoms and a continuum of waveguide modes modified in the
case of continuous couplings. As shown in Fig.~4(a) in the main
text, the two continuous couplings around $x_{1}=0$ and $x_{2}=d$
exhibit a common characteristic width $\Theta$, with which relevant
exponential distribution functions can be expressed as
$\nu_{1}(\varphi)=\frac{\sqrt{\Gamma}}{\Theta}e^{-\frac{2}{\Theta}|\varphi|}$
and
$\nu_{2}(\varphi)=\frac{\sqrt{\Gamma}}{\Theta}e^{-\frac{2}{\Theta}|\varphi-\phi|}$
that satisfy $\int d\varphi
\nu_{1,2}(\varphi)=\sqrt{\Gamma}$~\cite{continuum}. Here
$\varphi=\phi x/d\simeq\omega_{e}x/v_{g}$ describes the phase
accumulated from $x_{1}$ to $x$ by a propagating photon along the
waveguide and will become $\phi$ in the case of $x=x_{2}$, which has
been considered above for two discrete couplings. In this way, one
can immediately generalize the master equation~(2) in the main text
to the continuous-coupling case, where the modified constants are
given by
\begin{equation}
\begin{split}
\Gamma'&=\int_{-\infty}^{\infty}d\varphi \int_{-\infty}^{\infty} d\varphi'\nu_{1}(\varphi)\nu_{1}(\varphi')\text{cos}(\varphi-\varphi')\\
&=\frac{\Gamma}{\Theta^{2}}\int_{-\infty}^{\infty}d\varphi\int_{-\infty}^{\infty}  d\varphi' e^{-\frac{2}{\Theta}|\varphi|}e^{-\frac{2}{\Theta}|\varphi'|}\text{cos}(\varphi-\varphi')\\
&=\frac{\Gamma}{\Theta^{2}}\left[\int_{0}^{\infty}d\varphi\int_{0}^{\infty} d\varphi'e^{-\frac{2}{\Theta}\varphi}e^{-\frac{2}{\Theta}\varphi'}\text{cos}(\varphi-\varphi')
+\int_{0}^{\infty}d\varphi \int_{-\infty}^{0}d\varphi'e^{-\frac{2}{\Theta}\varphi}e^{\frac{2}{\Theta}\varphi'}\text{cos}(\varphi-\varphi')\right.\\
&\left.\quad\,+\int_{-\infty}^{0}d\varphi \int_{0}^{\infty}d\varphi'e^{\frac{2}{\Theta}\varphi}e^{-\frac{2}{\Theta}\varphi'}\text{cos}(\varphi-\varphi')
+\int_{-\infty}^{0}d\varphi\int_{-\infty}^{0} d\varphi'e^{\frac{2}{\Theta}\varphi}e^{\frac{2}{\Theta}\varphi'}\text{cos}(\varphi-\varphi')\right]\\
&=\frac{16\Gamma}{(\Theta^{2}+4)^{2}},
\end{split}
\tag{S13}
\label{Gap}
\end{equation}
\begin{equation}
\begin{split}
J_{ex}'&=\int_{-\infty}^{\infty}d\varphi\int_{-\infty}^{\infty}  d\varphi'\nu_{1}(\varphi)\nu_{2}(\varphi')\text{sin}|\varphi-\varphi'|\\
&=\frac{\Gamma}{\Theta^{2}}\int_{-\infty}^{\infty}d\varphi\int_{-\infty}^{\infty}  d\varphi' e^{-\frac{2}{\Theta}|\varphi|}e^{-\frac{2}{\Theta}|\varphi'-\phi|}\text{sin}|\varphi-\varphi'|\\
&=\frac{\Gamma}{\Theta^{2}}\left[2\int_{\phi}^{\infty}d\varphi\int_{\phi}^{\varphi} d\varphi'e^{-\frac{2}{\Theta}\varphi}e^{-\frac{2}{\Theta}(\varphi'-\phi)}\text{sin}(\varphi-\varphi')
+\int_{\phi}^{\infty}d\varphi\int_{0}^{\phi} d\varphi'e^{-\frac{2}{\Theta}\varphi}e^{\frac{2}{\Theta}(\varphi'-\phi)}\text{sin}(\varphi-\varphi') \right.\\
&\left.\quad\,+\int_{\phi}^{\infty}d\varphi \int_{-\infty}^{0}d\varphi'e^{-\frac{2}{\Theta}\varphi}e^{\frac{2}{\Theta}(\varphi'-\phi)}\text{sin}(\varphi-\varphi')
+\int_{0}^{\phi}d\varphi\int_{\phi}^{\infty} d\varphi'e^{-\frac{2}{\Theta}\varphi}e^{-\frac{2}{\Theta}(\varphi'-\phi)}\text{sin}(\varphi-\varphi')\right.\\
&\left.\quad\,+\int_{0}^{\phi}d\varphi \int_{0}^{\varphi}d\varphi'e^{-\frac{2}{\Theta}\varphi}e^{\frac{2}{\Theta}(\varphi'-\phi)}\text{sin}(\varphi-\varphi')
+\int_{0}^{\phi}d\varphi'\int_{0}^{\varphi'}d\varphi e^{-\frac{2}{\Theta}\varphi}e^{\frac{2}{\Theta}(\varphi'-\phi)}\text{sin}(\varphi'-\varphi)\right.\\
&\left.\quad\,+\int_{0}^{\phi}d\varphi \int_{-\infty}^{0}d\varphi'e^{-\frac{2}{\Theta}\varphi}e^{\frac{2}{\Theta}(\varphi'-\phi)}\text{sin}(\varphi-\varphi')
+\int_{-\infty}^{0}d\varphi \int_{\phi}^{\infty}d\varphi'e^{\frac{2}{\Theta}\varphi}e^{-\frac{2}{\Theta}(\varphi'-\phi)}\text{sin}(\varphi'-\varphi)\right.\\
&\left.\quad\,+\int_{-\infty}^{0}d\varphi \int_{0}^{\phi}d\varphi'e^{\frac{2}{\Theta}\varphi}e^{\frac{2}{\Theta}(\varphi'-\phi)}\text{sin}(\varphi'-\varphi)
+2\int_{-\infty}^{0}d\varphi\int_{-\infty}^{\phi}d\varphi'e^{\frac{2}{\Theta}\varphi}e^{\frac{2}{\Theta}(\varphi'-\phi)}\text{sin}(\varphi-\varphi')\right]\\
&=\frac{\Gamma}{(\Theta^{2}+4)^{2}}\left[(8\phi+2\Theta^{2}\phi+12\Theta+\Theta^{3})e^{-\frac{2\phi}{\Theta}}+16\text{sin}\phi\right],
\end{split}
\tag{S14}
\label{Uco}
\end{equation}
\begin{equation}
\begin{split}
\Gamma'_{ex}&=\int_{-\infty}^{\infty}\int_{-\infty}^{\infty} d\varphi d\varphi'\nu_{1}(\varphi)\nu_{2}(\varphi')\text{cos}(\varphi-\varphi')\\
&=\frac{\Gamma}{\Theta^{2}}\left[\int_{0}^{\infty}d\varphi\int_{\phi}^{\infty}  d\varphi' e^{-\frac{2}{\Theta}\varphi}e^{-\frac{2}{\Theta}(\varphi'-\phi)}\text{cos}(\varphi-\varphi')
+\int_{0}^{\infty}d\varphi\int_{-\infty}^{\phi}  d\varphi' e^{-\frac{2}{\Theta}\varphi}e^{\frac{2}{\Theta}(\varphi'-\phi)}\text{cos}(\varphi-\varphi')\right.\\
&\left.\quad\,+\int_{-\infty}^{0}d\varphi\int_{\phi}^{\infty}  d\varphi' e^{\frac{2}{\Theta}\varphi}e^{-\frac{2}{\Theta}(\varphi'-\phi)}\text{cos}(\varphi-\varphi')
+\int_{-\infty}^{0}d\varphi\int_{-\infty}^{\phi}  d\varphi' e^{\frac{2}{\Theta}\varphi}e^{\frac{2}{\Theta}(\varphi'-\phi)}\text{cos}(\varphi-\varphi')\right]\\
&=\frac{16\Gamma\text{cos}\phi}{(\Theta^{2}+4)^{2}},
\end{split}
\tag{S15}
\label{Gaco}
\end{equation}
with $\varphi'=\phi x'/v_{g}$ for a position $x'$ different from
$x$. Moreover, an extra coherent interaction term
$\sum_{j=1,2}J'\sigma_{+}^{j}\sigma_{-}^{j}$ has to be introduced in
$H_{at}$ with the effective interaction strength obtained as
\begin{equation}
\begin{split}
J'&=\int_{-\infty}^{\infty}d\varphi \int_{-\infty}^{\infty} d\varphi'\nu_{1}(\varphi)\nu_{1}(\varphi')\text{sin}|\varphi-\varphi'|\\
&=\frac{\Gamma}{\Theta^{2}}\int_{-\infty}^{\infty}d\varphi\int_{-\infty}^{\infty}  d\varphi' e^{-\frac{2}{\Theta}|\varphi|}e^{-\frac{2}{\Theta}|\varphi'|}\text{sin}|\varphi-\varphi'|\\
&=\frac{\Gamma}{\Theta^{2}}\left[2\int_{0}^{\infty}d\varphi\int_{0}^{\varphi} d\varphi'e^{-\frac{2}{\Theta}\varphi}e^{-\frac{2}{\Theta}\varphi'}\text{sin}(\varphi-\varphi')
+\int_{0}^{\infty}d\varphi \int_{-\infty}^{0}d\varphi'e^{-\frac{2}{\Theta}\varphi}e^{\frac{2}{\Theta}\varphi'}\text{sin}(\varphi-\varphi')\right.\\
&\left.\quad\,+\int_{-\infty}^{0}d\varphi \int_{0}^{\infty}d\varphi'e^{\frac{2}{\Theta}\varphi}e^{-\frac{2}{\Theta}\varphi'}\text{sin}(\varphi'-\varphi)
+2\int_{-\infty}^{0}d\varphi\int_{-\infty}^{\varphi} d\varphi'e^{\frac{2}{\Theta}\varphi}e^{\frac{2}{\Theta}\varphi'}\text{sin}(\varphi-\varphi')\right]\\
&=\frac{\Gamma\Theta(\Theta^{2}+12)}{(\Theta^{2}+4)^{2}}.
\end{split}
\tag{S16}
\label{Up}
\end{equation}
Note also that the expressions of $J'$ and $\Gamma'$ will remain
unchanged if we replace $\nu_{1}$ by $\nu_{2}$, i.e.,
\begin{equation}
\int_{-\infty}^{\infty}d\varphi \int_{-\infty}^{\infty} d\varphi'\nu_{1}(\varphi)\nu_{1}(\varphi')e^{i|\varphi-\varphi'|}=\int_{-\infty}^{\infty}d\varphi\int_{-\infty}^{\infty}  d\varphi'\nu_{2}(\varphi)\nu_{2}(\varphi')e^{i|\varphi-\varphi'|},
\tag{S17}
\label{exchange}
\end{equation}
implying that the two constants are identical for both Rydberg
atoms.

For the synthetic two-level giant atom, in a similar way,
$\Upsilon=(\Gamma+\Gamma_{ex}+iJ_{ex})\Omega_{c}^{2}/\Delta_{c}^{2}$
in the case of discrete couplings should be replaced by
$\Upsilon'=(\Gamma'+\Gamma'_{ex}+iJ'+iJ'_{ex})\Omega_{c}^{2}/\Delta_{c}^{2}$
in the case of continuous couplings.

Finally, we discuss how to control the coupling characteristic width
$\Theta$ in experiment by considering the radial size $\bar{r}$ of a
Rydberg atom. To be more specific, this characteristic width can be
estimated as
$\Theta=w\omega_{e}/v_{g}=2\sqrt{\bar{r}^{2}-h^{2}}\omega_{e}/v_{g}$
with $w$ being the overlap width between the electronic distribution
of the Rydberg atom and the evanescent field of the waveguide while
$h$ the distance from the Rydberg-atom nucleus to the
evanescent-field surface. Then it is viable to attain
$\Theta=5\pi/2$ with $h\simeq449\,\text{nm}$ since we have
$\bar{r}\simeq583\,\text{nm}$ for
$|r_{1,2}\rangle=|75P_{3/2}\rangle$~\cite{website}.

\end{document}